
\magnification 1200
\baselineskip=18pt
\nopagenumbers
\overfullrule=0pt
\def\rs{\vbox{\hbox{\raise1.6mm\hbox{$>$}}
\kern-18pt\hbox{\lower1.6mm\hbox{$\sim$}}}}
\def\ls{\vbox{\hbox{\raise1.6mm\hbox{$<$}}
\kern-18pt\hbox{\lower1.6mm\hbox{$\sim$}}}}
\centerline{\bf SUPERGRAVITY MODELS}
\bigskip
\centerline{R. Arnowitt\footnote*{Speaker}$^a$ and Pran Nath$^b$}

\centerline{$^a$Center for Theoretical Physics,
Department of Physics}
\centerline{Texas A\&M University, College Station, TX  77843-4242}
\centerline{$^b$Department of Physics, Northeastern University, Boston, MA
02115}
\medskip
\centerline{ABSTRACT}
\smallskip

Theoretical and experimental motivations behind supergravity grand unified
models are described.  The basic ideas of supergravity, and the origin of the
soft breaking terms are reviewed.  Effects of GUT thresholds and predictions
arising from models possessing proton decay are discussed.  Speculations as to
which aspects of the Standard Model might be explained by supergravity models
and which may require Planck scale physics to understand are mentioned.
\medskip

\noindent
1.  INTRODUCTION
\smallskip

Supergravity has become the main vehicle for efforts to construct grand
unified models.  There are now both experimental and theoretical reasons for
examining the consequences of such models.  On the experimental side, there is
the well known
fact that measurements of $\alpha_1\equiv(5/3)~\alpha_Y$, $\alpha_2$ and
$\alpha
_3$ at mass
scale Q = $M_Z$ (where $\alpha_Y$ is the hypercharge coupling constant) allows
a test of
whether these three couplings of the Standard  Model unify at some high scale Q
= $M_G$.  What
is found [1] is that unification does not occur with the Standard Model (SM)
mas
s spectrum but
unification does appear to occur with the supersymmetrized Standard Model with
one pair of
Higgs doublets.  Thus using the two loop renormalization group equations and
making the
approximation of neglecting both mass splitting of the supersymmetry (SUSY)
spectrum and mass
splitting of the GUT mass spectrum, one finds that

$$M_G=10^{16.19\pm 0.34}~{\rm GeV};\quad M_S=10^{2.37 \pm 1.0}~{\rm GeV}$$
$$\alpha_G^{-1} = 25.4 \pm 1.7\eqno(1)$$

\noindent
where $M_S$ is the common SUSY mass, and $\alpha_G$ is the gauge coupling
constant at the unification scale $M_G$.  (The errors in Eq. (1) are due to
the errors in $\alpha_3 (M_Z)$ and we use $\alpha_3 (M_Z) = 0.118 \pm 0.007$
[2].)
\smallskip
There are several points worth noting about the above result:

\item{(i)}  Unification occcurs only for the choice
$\alpha_1\equiv (5/3)\alpha_Y$,
which states the way in which the hypercharge is embedded into the GUT group
G.  Thus unification is not completely a property of the low energy particle
spectrum, but depends also on the nature of the high energy group G.

\item{(ii)}  Unification is indeed obtained by adjusting the parameter $M_S$.
However, the significant point is that $M_G$ and $M_S$ come out at values that
are physically acceptable, i.e. $M_G$ is sufficiently large to inhibit proton
decay, and $M_S$ is in the correct mass region for the SUSY particles to solve
the gauge hierarchy problem discussed below.  (Thus $M{_S}\cong 10^{2.5}
\cong 300$ GeV.)

\item{(iii)}  Acceptable unification occurs only with one pair of Higgs
doublets.
With more Higgs doublets, $M_G$ is so small that proton decay would already
have been observed, and $M_S$ is so large that the hierarchy problem remains.
\smallskip
Of course, we have no real knowledge of what the particle spectrum is above the
electroweak scale.  There may be additional particles at higher energies which
delay or prevent grand unification from occurring.  However, the simplest and
most natural implication of the above result is that grand unification occurs
at scale $M_G$, and the particle spectrum between $M_Z$ and $M_G$ is the
supersymmetrized Standard Model with one pair of Higgs doublets.
\smallskip
There are also several theoretical arguments supporting the building of
supersymmetric particle models.  From the high energy side, string theory
implies the validity of N = 1 supergravity as an effective field theory below
the Planck scale, $M_{P\ell} = (1/8 \pi G_N)^{1/2}$ where $G_N$ is the
Newtonian constant $(M_{P\ell} = 2.4 \times 10^{18}~{\rm GeV})$.  Note,
however,
 that
$M_G$/M$_{P\ell} \approx 10^{-2}$ and so the GUT theory is moderately
isolated from Planck scale physics.  However, we do not expect it to be a
precisely accurate theory as it may possess (1-10) \% corrections from ``Planck
slop" terms (non-renormalizable terms scaled by powers of 1/M$_{P\ell}$).
\smallskip
{}From the low energy electroweak scale, supersymmetry offers a solution to the
well-known gauge hierarchy problem.  Thus in the Standard Model, the loop
corrections to the Higgs mass $m_H$ (Fig.1) is quadratically divergent:
$$m_H^2 = m_0^2 + c({\tilde\alpha}/4\pi)\Lambda^2\eqno(2)$$
\noindent
where $m_0$ is the bare mass, ${\tilde\alpha}$ is a coupling constant, $c$ is a
numeric and $\Lambda$ is the cut-off.  If one takes the bare Lagrangian as
fundamental, then the existence of the divergence implies that the theory is
valid at energies below $\Lambda$, and $\Lambda$ is the scale of new physics
which intervenes to converge the integral.  How large can $\Lambda$ be, i.e. at
what scale does new physics enter?  Now m$_H$ sets the electroweak scale.
However, as $\Lambda$ gets large, m$_H$ and eventually other particle masses
all get close to the large scale $\Lambda$.  This is the ``gauge hierarchy"
problem which states that it is not possible to maintain a hierarchy of masses,
some small at the electroweak scale and some large (e.g. at $M_G$ or
M$_{P\ell}$ scale).  An alternate way of thinking of this problem is to try to
choose m$_0^2$ to cancel the large $\Lambda^2$ term.  However, for
$\Lambda \approx M_G \approx 10^{15}~{\rm GeV}$, this requires fine tuning
$m_0^2$ to 24 decimal places (!) and trouble begins already for $\Lambda\rs
$ 1 TeV.  This alternate view of the problem is known as the ``fine
tuning" problem.  Of course, the same difficulties enter with the other
divergences of relativisitic quantum field theory.  However, these only grow
logarithmically with $\Lambda$, and so hierarchy difficulties only set in at
the Planck scale where we already know new physics must occur.

{}~\vskip 2.15truein
\noindent
 Fig. 1  One loop correction to Higgs self mass from Higgs
coupling to quarks.
\noindent
\smallskip
Solutions to the gauge hierarchy problem fall into two categories:  either one
assumes the Higgs is composite (e.g. as  in technicolor or $t\bar t$ condensate
models) and hence dissociates at scale $\Lambda$, or one assumes a symmetry
exists to cancel the quadratic divergences.  The latter possibility is
supersymmetry where the Bose-Fermi symmetry causes this cancellation.  For
perfect supersymmetry, the two diagrams of Fig. 2 precisely cancel.  If
supersymmetry is broken by lifting the squark-quark degeneracy then the
quadratic divergence still cancels leaving an underlying logarithmic
divergence:

$$\Lambda^2\rightarrow (m_{\tilde q}^2  - m_q^2)\ell n(\Lambda^2/m_{\tilde
q}^2)
\eqno(3)$$

\noindent
Thus to avoid fine tuning we need $m_{\tilde q} \rs~1~{\rm TeV}$, i.e. $M_S \ls
1~{\rm TeV}$ and
the new SUSY particles lie within the range for detection by current and
planned
 accelerators.
In fact, for a wide class of models it has been shown that $m_h \ls 146~{\rm
GeV
}$ [3] (and
usually $m_h \ls 120~ {\rm GeV}$) which would make the light Higgs accessible
to
LEP200 or its upgrades.
\vskip 2.15truein

Fig. 2.  Higgs one loop corrections in supersymmetric models.  $\tilde q$ are
spin zero squarks.
\bigskip
\noindent
2.  TRIVIALITY BOUND:  AN ALTERNATE VIEW
\medskip
The analysis given above takes the viewpoint that the bare Lagrangian is the
fundamental quantity.  However, the Standard Model is a renormalizable field
theory.  One can therefore pre-renormalize it (by introducing counter terms)
and deal only with finite renormalized Green's functions.  Masses and coupling
constants can then be defined by these Green's functions at fixed momenta e.g.
for the renormalized Higgs propagator $\Delta ^{(R)}_H (q^2)$ one may
define the Higgs mass parameter $m_H$ by $m_H^2$ = [$\Delta ^{(R)}_H
(0)$]$^{-1}$.  The $Z_2$ rescaling of $\Delta ^{(R)}_H$ can be defined by
the condition $[\partial(\Delta ^{(R)}_H)^{-1}/\partial q^2]_{q^2=0} =1$.
Similarly, the $\lambda\phi^4$ coupling constant may be defined from the
renormalized 4-point vertex ${\Gamma ^{(R)}_4(p_1,p_2,p_3)}$ by ${\lambda =
\Gamma ^{(R)}_4(0,0,0)}$.
\smallskip
In the tree approximation, one has $V_H = -m^2\phi^+\phi+\lambda(\phi^+\phi)^2$
with $m^2$, $\lambda > 0$, and defining $\langle\phi\rangle\equiv
v/\sqrt{2}$ one finds $\langle\phi \rangle^2 = m^2/2\lambda$ and the Higgs mass
to
be $m_H^2 = 2m^2$.  Since $M_W = g_2 v/2$ (and hence $v \cong 247~{\rm GeV}$)
one may write

$$M_W = {g_2\over 2\sqrt{2\lambda}} m_H\eqno(4)$$

\noindent
showing that the Higgs mass scales electroweak physics, and also that

$$\lambda = {g_2^2\over 8}{m_H^2\over M_W^2}\eqno(5)$$

\noindent
If one takes now the alternate viewpoint that the renormalized field theory is
thefundamental theory, one never sees a quadratic divergence (or any other
divergence).  Thus the theory has no problems unless it is internally
inconsistent (under which circumstances it would self-destruct).  This actually
happens, as the theory develops a Landau pole.  Letting $\lambda(Q)$ be the
running coupling constant, one finds, in the approximation of keeping only the
Higgs self-couplings of $V_H$, the result

$$\lambda(Q) = {\lambda(M_W)\over 1 -{3\lambda(M_W)\over 4\pi^2} \ell
n(Q^2/M_W^2)}\eqno(6)$$

\noindent
where $\lambda(M_W)$ is the low energy value given approximately by Eq. (5).
A pole occurs in Eq. (6) at scale $Q_0$ where the denominator vanishes.  The
theory breaks down at $Q\approx Q_0$ and so $Q_0$ must be a scale where new
physics sets in.  Using Eq. (5) one finds for this scale

$${3\over 4\pi}\alpha_2 {m_H^2\over M_W^2} \ell n(Q_0/M_W) = 1\eqno(7)$$

\noindent
In this viewpoint, the scale of new physics is determined by the {\it
experimental} value of the Higgs mass, and the lighter the Higgs mass the
larger

$Q_0$ is.  For example, if $m_H = 146 ~{\rm GeV}$ one finds $Q_0 \cong
M_{P\ell}
$
while if $m_H = 500~ {\rm GeV}$ then $Q_0 \cong$ 2 TeV. Thus, if the Higgs is
light,
the Standard Model could hold all the way up to the Planck scale.  If the Higgs
is heavy, the Standard Model must break down in the TeV range implying an upper
limit on $m_H$.  (Of course, the argument does not exclude new physics from
arising before $Q_0$ from some other cause, but only that $Q_0$ is an upper
bound on the validity of the SM.)
\smallskip
The analysis given here can be extended to include gauge and Yukawa couplings,
and has been performed using lattice gauge theory (as the theory becomes
non-perturbative near the Landau pole).  The above results remain qualitatively
correct.  (See, e.g. Ref [4].)  Which viewpoint, the previous discussion of the
gauge hierarchy problem or the Landau pole problem, determines the scale where
new physics must arise depends on whether one believes the bare or renormalized
theory is fundamental.  In this discussion we take the gauge hierarchy problem
as fundamental, and discuss the consequences of the supersymmetric solution to
this difficulty.
\medskip
\noindent
3.  SUSY BASICS
\smallskip
In supersymmetry, multiplets must have an equal number of Fermi and Bose
helicity states.  To build a supersymmetrized Standard Model, one needs two
types of massless multiplets, chiral multiplets and vector multiplets.
\smallskip
\noindent
Chiral Multiplets: $ (z(x)$, $\chi(x))$
\smallskip
Here $z(x)$ is a complex scalar field (s=0) and $\chi(x)$ is a left-handed (L)
Weyl spinor (s = 1/2).  Thus $\chi(x)$ can be used to represent quarks and
leptons and also the spin 1/2 Higgsino partners of the Higgs boson, while the
$z(x)$ can be used to represent the Higgs boson and the spin 0 squarks and
slepton partners of the quarks and leptons
\smallskip
\noindent
Vector Multiplets: $ (V^\mu (x), \lambda (x))$
\smallskip
Here the $V^\mu(x)$ are real vector fields (s = 1) representing the gauge
bosons, and $\lambda (x)$ are Majorana spinors (s = 1/2) representing the
gaugino partners.
\smallskip
The Higgs doublets must come in pairs in supersymmetry to cancel anomalies.
The minimal number is just two:

$$H_1 = (H_1^0, H_1^-); H_2 = (H_2^+, H_2^0)\eqno(8)$$

\noindent
The dynamics of global supersymmetry consists of gauge interactions
(supersymmetrized) and
Yukawa interactions governed by the superpotential $W$.  In general, $W(z_a)$
is
 a holomorphic
function of the scalar fields $z_a$ and
hence independent of the $z_a^{\dagger}$.  For
renormalizable interactions, $W$ is at most cubic in the fields.  Thus the most
general
renormalizable $SU(3)_C \times SU(2)_L \times U(1)_Y$ and $R$ parity
invariant form for $W$ is

$$\eqalign{W = \mu H^{\alpha}_1 H_{2\alpha} &+ [\lambda ^{(u)}_{ij}q^{\alpha}
_i H_{2\alpha} u ^C_j + \lambda ^{(d)}_{ij}q^{\alpha}_i H_{1\alpha}d ^C_j\cr
&+\lambda ^{(\ell )}_{ij} \ell^{\alpha}_i H_{1\alpha}e ^C_j]\cr}\eqno(9)$$

\noindent
Here $i, j = 1,2,3$ are generation indices, $\alpha = 1,2$ is the $SU(2)_L$
 index
$(H_\alpha = \varepsilon_{\alpha\beta}H^{\beta}$, $\varepsilon_{\alpha\beta} =
-
\varepsilon_{\beta\alpha}$, $\varepsilon_{12} = +1)$, $C$ = charge conjugate,
$\lambda^{(u,d,\ell)}_{ij}$ are Yukawa coupling constants
and $\mu$ is a mass scaling
the Higgs mixing term.  Note that the gauge invariant $u$-quark interaction
requires the $H_{2\alpha}$ Higgs doublet to appear, since
$H_{1\alpha}^\dagger$ cannot enter as $W$ is holomorphic.  Thus the existance
of two Higgs doublets is also necessary to obtain mass growth of both the up
and down quarks.
\smallskip
The supersymmetry invariant dynamics can be described by an effective potential

\noindent
$$V = \sum_a\mid {\partial W\over \partial Z_a}\mid ^2 + V_D;~ V_D = {1\over
2}~
 g^2_i D_{ir}D_{ir}\eqno(10)$$

\noindent
(where $g_i$  are the $SU(3) \times SU(2) \times U(1)$ coupling constants,
$D_{ir} =
z_a^{\dagger} (T^{ir})_{ab}z_b$, $T_{ab}^{ir} = $group generators), and
fermionic interactions

$${\cal L}_Y = - {1\over 2}\sum_{a,b}({\bar \chi}^{aC}{\partial^2W\over\partial
z_a\partial z_b} \chi^b + h.c.)\eqno(11)$$

\noindent
and

$${\cal L}_\lambda = -i\sqrt{2}~\sum g_i {\bar \lambda}^{ir} z^{\dagger}_a
(T^{ir})^a_b \chi_b + h.c.\eqno(12)$$

\noindent
Note that $V_D$ plays the role of the $\lambda (\phi^+\phi)^2$ term in the SM,
but with $\lambda$ replaced by the gauge coupling constants $g_i$.  It is this
that allows SUSY predictions of Higgs mass bounds since the $g_i$ are known.
\smallskip
After $SU(2) \times U(1)$ breaking, the Higgsinos and $SU(2) \times U(1)$
 gauginos mix.
There result 32 new SUSY particles: (i)  12 squarks (s=0, complex):  ${\tilde
q}
_i = ({\tilde
u}_{iL}, {\tilde d} _{iL})$; ${\tilde u}_{iR}$, ${\tilde d}_{iR}$;
(ii) 9 sleptons (s=0,
complex) ${\tilde \ell}_i = {\tilde \nu}_{iL}, {\tilde e}_{iL}); {\tilde
e}_{iR}
$; (iii) 1
gluino ($s={1\over 2}$, Majorana) $\lambda^a$, $a = 1\ldots 8 = SU(3)_C$ index;
(iv) 2 Winos
(Charginos) $(s={1\over 2}$, Dirac).  ${\tilde W}_i$, $i = 1,2$, $m_i <m_j$ for
$i < j$; (v) 4
Zinos (Neutralinos) (s = 1/2, Marjorana) ${\tilde Z}_i$, $i = 1\ldots 4$, $m_i
<
 m_j$
for $i < j$; and (vi) 4 Higgs (s = 0) $h^0$, $H^0$ real CP even; $A^0$ real CP
odd; $H^{\pm}$ charged.
\smallskip
The $h^0$ is the particle which most resembles the SM Higgs.
\medskip
\noindent
4.  SUPERGRAVITY BASICS
\smallskip

The global SUSY models discussed in the previous section possesses one serious
drawback:  it is not possible to achieve a phenomenologically satisfactory
spontaneous breaking of supersymmetry.  There are a number of reasons for
this.  Most obvious is that the breaking of a global symmetry implies the
existence of a massless Goldstone particle, in this case a spin 1/2 particle
(the Goldstino), and no candidate exists experimentally.  (The neutrino
interactions do not obey the correct threshold theorems.[9a])  An obvious
solution
to this difficulty is to promote supersymmetry to a local symmetry.  The gauge
particle is then spin 3/2 (the gravitino) and upon breaking of supersymmetry
it absorbs the spin 1/2 Goldstino to become massive.  However, supersymmetry
requires that the gravitino be embedded in a massless multiplet
$(g_{\mu\nu}(x)$; $\psi_\mu(x))$.  Here $g_{\mu\nu}(x)$ is a massless spin 2
field i.e. one is led to supergravity theory [5] where gravity is automatically
included.
\smallskip
The coupling of supergravity to chiral and vector matter multiplets depends
upon the following functions [6-9]:  the superpotential $W(z_a$), the K\"ahler
potential $d(z_z, z_a^{\dagger})$, and the gauge kinetic function
$f_{\alpha\beta} (z_a, z_a^{\dagger})$ where $\alpha,\beta$ are gauge indices.
Actually, $W$ and $d$ enter only in the combination ${\cal G}$ = -$\kappa ^2d -
\ell n~
[\kappa ^6WW^{\dagger}]$ where $\kappa \equiv 1/M_{P\ell}$.  We will assume in
the following that $d$ and $f_{\alpha\beta}$ can be expanded in powers of
fields
with the higher non-linear terms scaled by $\kappa$:

$$d(z_a,z_a^{\dagger}) = c_b^az_az_b^{\dagger} +~(a^{ab}z_az_b + h.c.)
+ \kappa c^{ab}_c z_az_bz_c^{\dagger}+\cdots\eqno(13)$$

$$f_{\alpha\beta}(z_az_a^{\dagger}) = c_{\alpha\beta} + \kappa
(c^a_{\alpha\beta} z_a + h.c.)+\cdots\eqno(14)$$

\noindent
Supersymmetry breaking can occur at the tree level [10] or via condensates
[11] due to supergravity interactions.  The simplest example is to choose $W =
m^2(z + B)$, $d = z_az_a^{\dagger}$ and minimizing the effective potential one
finds $\langle z \rangle = \pm \kappa^{-1}(\sqrt{2} - \sqrt{6}) =
O(M_{P\ell})$.
  (One may
further chose B to fine tune the cosmological constant to zero.)  The quantity
$M_S = O(\langle\kappa^2W \rangle) \sim \kappa m^2$ will turn out to scale
the SUSY masses.
\smallskip
The full supergravity dynamics is quite complicated.  (For a discussion see
Refs. [8,12].)  We list here some of the important terms.  The effective
potential is given by

$$V = e^{\kappa d}[(g^{-1})_b^a({\partial W\over \partial z_a} + \kappa^2
d_aW)({\partial W\over \partial z_b} + \kappa^2 d_bW)^{\dagger} - 3\kappa^2
\mid W\mid^2] + V_D\eqno(15)$$

\noindent
where

$$V_D = {1\over 2}g^2Re(f^{-1})~_{\alpha\beta}D_{\alpha}D_{\beta};~ D_{\alpha}
=
d^a(T^{\alpha})_{ab}z_b\eqno(16)$$

\noindent
$d_a = \partial d/\partial z_a^{\dagger}$, $d^a= \partial d/\partial z_a$ and
$g_b^a =
\partial ^2 d/\partial z_b\partial z_a^{\dagger}$, and $\alpha,\beta$ are gauge
indices.
Thus there are $\kappa = 1/M_{ P\ell}$ corrections to Eq. (10).  The scalar
field kinetic energy is -${1\over 2} g^a_b(D_{\mu}z_b)(D_{\mu}z_a)^{\dagger}$
(where
$D_{\mu}$ is the covariant derivative).
{}From Eq.(13), $g^a_b = c^a_b + O (\kappa)$ and so
diagonalizing c$^a_b$ brings the scalar kinetic energy into canonical form.
The
$O(\kappa)$ correction are non-renormalizable corrections scaled by
$1/M_{P\ell}
$, and
presumably small below the GUT scale.  The gauge field kinetic energy is
$-{1\over 4}$(Re f$_{\alpha\beta}) F^{\alpha}_{\mu\nu} F ^{\mu\nu\beta}$ and
from Eq.(14) one sees
that one obtains the canonical kinetic energy plus possible 1/M$_{P\ell}$
corrections.  Finally, there is a gaugino term

$$[e^{\kappa^2d} \kappa^2\mid W\mid(g^{-1})^a_b d^b f^{\dagger}_{\alpha\beta
a}]
 {\bar
\lambda}^{\alpha}\lambda^{\beta}\eqno(17)$$

\noindent
where $f_{\alpha\beta a} \equiv \partial f_{\alpha\beta}/\partial z_a$.

To see the effects of supersymmetry breaking we consider the simple tree model
discussed above where $W = m^2 (z+B)$ and $d= z_az_a^{\dagger}$.  From Eq. (15)
one has
the term

$$(\kappa^2d_aW)(\kappa^2d_aW)^{\dagger}\rightarrow(\kappa^2~\langle
 W~\rangle)^2z_az_a^{\dagger}\eqno(18)$$

\noindent
Thus each scalar field grows a universal mass $m^2_0$ =($\kappa^2 \langle W
\rangle)^2 = O(M^2_S$).  From Eqs. (17) and (14) for the case $z_a=z$, a
universal gaugino mass, $m_{1/2}$, forms of size

$$\langle\kappa^2\mid W\mid(\partial d/\partial z) f^{\dagger}_{\alpha\beta z}
\rangle =\langle\kappa^2\mid W\mid z^{\dagger}\kappa c^z_{\alpha\beta}
 \rangle\eqno(19)$$

\noindent
and so $m_{1/2}$ = $O(M_S)$.  Further, by transforming the second term of Eq.
(13) from the K\"ahler potential to the superpotential by a K\"ahler
transformation,
a Higgs mixing parameter $\mu_0$ forms where

$$\mu_0 = \langle\kappa^2W\rangle a^{H_1H_2} = O(M_S)\eqno(20)$$

\noindent
Finally, two additional supersymmetry breaking structures arise from Eq. (15)
when the matter parts of the superpotential are included:

$$A_0W^{(3)} + B_0W^{(2)}\eqno(21)$$

\noindent
where $W^{(2,3)}$ are the (quadratic, cubic) parts of the matter
superpotential.

One finds here also that

$$A_0,B_0 = O(M_S)\eqno(22)$$

\noindent
so that supersymmetry breaking gives rise to four soft breaking terms scaled by
$m_0$, $m_{1/2}$, $A_0$, $B_0$ and a supersymmetric Higgs mixing parameter
$\mu_0$.  All these parameters are $O(M_S)$.
\medskip
\noindent
5.  RADIATIVE BREAKING
\smallskip
A remarkable feature of supergravity GUT models is that they offer a natural
explanation of $SU(2) \times U(1)$ beaking via radiative corrections [13].
In theStandard Model, $SU(2) \times U(1)$ breaking is {\it accomodated}
by the device of
having a negative $(mass)^2$ for the Higgs.  However, no explanation is given
as to why this choice should be made.  We saw in Eq. (18), that supersymmetry
breaking gives rise to a universal positive $(mass)^2$, $m_0^2 > 0$, at the
scale $Q = M_G$.  One may now run the renormalization group equations (RGE)
down to the electroweak scale.  As shown schematically in Fig. 3, $m_{H_2}^2$
bends downward and eventually turns negative (due to the t-quark Yukawa
couplings) signaling the breaking of $SU(2) \times U(1)$.

\vskip 3.0truein
\noindent

Fig. 3.  Schematic diagram of running masses showing that
$m_H^2$ turns negative at the electroweak scale due to the heavy top
interactions.
\smallskip
To see the above more quantitatively, the renormalizable Higgs interactions
from Eq. (15), have the form

$$\eqalign{V_H&=m_1^2(t)\mid H_1\mid^2+m_2^2(t)\mid H_2\mid^2-m_3^2(t)(H_1H_2 +
h.c.)\cr
&+{1\over 8}[g^2_2(t) + g_Y^2(t)][\mid H_1\mid^2-\mid H_2\mid^2]^2 +
\Delta V_1\cr}\eqno(23)$$

\noindent
where $\Delta V_1$ is the one loop addition, and all parameters are running
with respect to the variable $t = \ell n[M_G^2/Q^2]$.  In Eq. (23), the
masses are defined by $m_i^2(t) = m_{Hi}^2(t) + \mu^2(t)$, $i = 1,2$ and
$m_3^2(t) = -B(t) \mu(t)$ subject to the boundary conditions at $Q = M_G$
of $m_i^2 (0) = m_0^2 + \mu_0^2, m_3^2(t) = -B_0\mu_0$.  Minimizing the
effective potential, $\partial V_H/\partial v_i=0$,
$v_i\equiv \langle H_i\rangle $ one finds

$${1\over 2}M_Z^2 = {\mu_1^2 -\mu_2^2{\rm tan}^2\beta\over {\rm
tan}^2\beta-1};~
 {\rm sin}
2\beta = {2m_3^2\over \mu_1^2 + \mu_2^2}\eqno(24)$$

\noindent
where $\mu_i^2 = m_i^2 + \Sigma_i$ and ${\rm tan} \beta\equiv v_2/v_1$.
($\Sigma_i$
are the one loop corrections.)  The RGE allows one to express all the
parameters in Eq. (24) in terms of the GUT scale constants $m_0, m_{1/2}, A_0,
B_0$ and $\mu_0$.  One may use Eq. (24) to eliminate $B_0$ and $\mu_0^2$ in
terms of the remaining constants and tan $\beta$.  Thus one can express all 32
SUSY masses in terms of the four parameters $m_0, m_{1/2}, A_0$ and tan
$\beta$ and the as yet undetermined top quark mass $m_t$.  Since the sign of
$\mu_0$ is not determined there are two branches:  $\mu_0 < 0$ and
$\mu_0 > 0$.  It is interesting to ask under what conditions will a
satisfactory electroweak breaking occur.  Three necessary conditions are:
(i)  Not all the soft breaking parameters, $m_0, m_{1/2}$, $A_0$, and $B_0$ can
be
zero; (ii) $\mu_0$ must be non-zero; (iii) $m_t$ must be large $(m_t >\rs
90 ~{\rm GeV})$.
Thus in a real sense, item (i) implies that supersymmetry breaking
triggers electroweak breaking, and from (iii) the existence of electroweak
breaking predicts that the top must be heavy.
\medskip
\noindent
6.  SIMPLE GUT MODEL
\smallskip

In Sec. 1, grand unification was discussed neglecting, however, the
existence of possible GUT states which would produce threshold corrections in
the vicinity of $M_G$.  In order to see the size of these effects, we examine
here a simple $SU(5)$ model first proposed within the framework of global
supersymmetry [14].  GUT physics here is characterized by the superpotential

$$W_G = \lambda_1[{1\over 3}
Tr\Sigma^3+{1\over 2}M Tr\Sigma^2]+\lambda_2H^Y[\Sigma
^X_Y + 3M'\delta^X_Y]{\bar H}_X)\eqno(25)$$

\noindent
Here $\Sigma^X_Y (X,Y  = 1\ldots 5)$ is a 24 of $SU(5)$, while $H^Y$ and ${\bar
H}_X$
are a 5 and ${\bar 5}$ of $SU(5)$.  The $SU(2)$ doublets of $H^X$ and
${\bar H}_X$ are just the
$H_1$ and $H_2$ doublets of low energy theory.
They are kept light by the choice $M = M'$,
(which we will make here) though more natural ways of keeping the
Higgs doublets light exist [15].  Upon minimizing the effective potential
$\Sigma^X_Y$ grows the VEV

$${\rm diag} \langle\Sigma^X_Y\rangle = M(2,2,2,-3,-3)\eqno(26)$$

\noindent
breaking $SU(5)$ to the SM.  We have then that $M = O(M_G)$.  The states
that become superheavy are the color triplets of $H^X$ and ${\bar H}_X$
transforming like (3,1) and (${\bar 3},1$) under $SU(3)_C \times SU(2)_L$ with
 mass
$M_{H_3} = 5\lambda_2M$, massive vector multiplets, transforming as (3,2) and
($
{\bar3},2$) with mass $M_V = 5 \sqrt{2} gM~ (\alpha_G\equiv g^2/4\pi)$
and the superheavy
components of $\Sigma^X_Y$ transforming as (8,1), (1,3) and (1,1)
with masses $M^8_\Sigma =
5\lambda_1 M/2 = M^3_\Sigma~ {\rm and}~ M^0_\Sigma = \lambda_1 M/2$.  This
model
 has been
considered previously [16] (though with inaccurate arguments).
\smallskip
We limit here $\lambda _{1,2} < 2$ (so that one stays within the perturbative
domain) and also require $\lambda_{1,2} > 0.01$ (so that the superheavy spectra
stay in the GUT range).  When thresholds are ignored, the RGE can be used to
predict a value for $\alpha_3(M_Z)$.
 With thresholds, one gets instead a correlation between
$\alpha_3(M_Z)$ and $M_{H_3}$. As seen in Fig. 4 [17], one obtains an
upperbound
 of
$\alpha_3(M_Z) < 0.135$.  Since current proton decay data requires $M_{H_3}\rs
1
 \times
10^{16}~{\rm GeV}$, one also gets a lower bound of $\alpha_3(M_Z) > 0.114$.
These are
consistent with the current experimental bounds of $\alpha_3(M_Z) = 0.118 \pm
0.
007$.  For the
1$\sigma$ upper limit of $\alpha_3(M_Z) = 0.125$, one finds $M_{H_3} < 2 \times
10^{17}{\rm GeV}$, and so $M_{H_3}$ is always below the Planck scale [18].
Thus
 the
model gives generally reasonable results.  Measurements of the SUSY particle
masses would determine $M_S$ which corresponds in Fig. 4 to a line in between
the $M_S = 1~{\rm TeV}$ and $M_S = 30~ {\rm GeV}$ bounding lines.  That, plus
an
 accurate
measurement of $\alpha_3(M_Z)$, would determine a point within the
quadrilateral and hence fix $M_{H_3}$.  Thus accurate low energy measurements
would allow a prediction of the proton decay rate for $p \rightarrow {\bar \nu}
K^+$, i.e. the model can also be experimentally tested!
{}~\vskip 3.5truein
\noindent
Fig. 4.  Grand unification constraints for the GUT model of Eq.
(25).  Grand unification correlates $\alpha_3(M_Z)$ with $M_{H_3}$.  The
allowed
region is within the solid quadrilateral.

\medskip
\noindent
7.  GUT PHYSICS OR PLANCK PHYSICS?
\smallskip

Supergravity GUT models do not represent a fundamental theory, but rather an
effective theory valid at energies below $M_G$.  One may ask what aspects of
the theory can be understood at the GUT level, and what requires higher scale
physics, presumably unknown Planck scale physics, to understand.  We list here
a few speculations.

\item{(i)}  Unification of gauge couplings.
 This is presumably GUT physics since it
depends on the particle spectrum below $M_G$ and on the grand unification group
$G$ which holds above $M_G$.

\item{(ii)}  Quark/lepton masses, KM matrix elements, Yukawa coupling constants
are
presumably Planck scale physics (e.g. as in string theory) except for possible
symmetry constraints that the GUT group G may impose.

\item{(iii)}  Nature of supersymmetry breaking.
The structure of the hidden sector
where supersymmetry breaking takes place is presumably Planck physics.
However,
it can be parameterized at the GUT scale in terms of five parameters $m_0$,
$m_{1/2}$, $A_0$, $B_0$ and $\mu_0$.

\item{(iv)}  Squark/slepton masses and widths.
This is GUT physics, once the five
hidden sector parameters are chosen.

\item{(v)}  Electroweak breaking.  This is GUT physics, once the hidden sector
parameters are chosen.

\item{(vi)}  Proton stability.  GUT physics depending on the interactions which
break $G$ to the SM group.
\smallskip
We see from the above, that while supergravity grand unified models add
significantly to our understanding of low energy physics, there are a number of
areas, notably in the Yukawa couplings and in the structure of the hidden
sector, where it offers no new insights.  For these one must make a
phenomenological treatment.

\medskip
\noindent
8.  PROTON DECAY
\smallskip

There are two main modes of proton decay in GUT models:  $p\rightarrow e^+ +
\pi^0$ and $p\rightarrow {\bar \nu} + K^+$.  The former can occur in both
SUSY and non-SUSY grand unification, and generally will occur for any model
whose grand unification group $G$ possesses $SU(5)$ as a subgroup.
The latter is a
specifically supersymmetric mode.  Thus the observation of $p\rightarrow {\bar
\nu}K^+$ would be a strong indication of the validity of supergravity
grand unification.  This decay can also occur when $G$ possesses an $SU(5)$
subgroup and if the light matter below $M_G$ is embedded in the usual way in 10
and $\bar 5$ representations of the $SU(5)$ subgroup.
 However, it is possible to
construct a complicated Higgs sector where one fine tunes the $p\rightarrow
{\bar \nu}K$ amplitude to zero and still maintains only too light Higgs
doublets
below $M_G$.  However, such models appear somewhat artificial, and the $p
\rightarrow {\bar \nu}K$ decay mode is generally expected to arise, though it
can be evaded.

\item{(i)}{  $p\rightarrow e^+\pi^0$.
 This mode proceeds as in non-SUSY GUTs through
the superheavy vector bosons of mass $M_V = O(M_G)$.  For SUSY models one
 has [19]

$$\tau(p\rightarrow e^+\pi^0) = 10^{31\pm 1}({M_V\over 6 \times 10^{14}~{\rm
GeV
}})^4
{\rm yr}\eqno(27)$$
The current experimental bound is [20] $\tau(p\rightarrow e^+\pi^0) > 5.5
\times
10^{32}$ yr $(90\%~{\rm CL})$.  Super Kamiokande expects to be sensitive up to
a
lifetime $\tau(p\rightarrow e^+\pi^0) < 1 \times 10 ^{34}$ yr [21].  From
 Eq. (27)
this would require $M_V \ls 6 \times 10^{15}~{\rm GeV}$ for the decay mode to
be
observable.}

\item{(ii)}  $p\rightarrow {\bar \nu} K^+$.  For the models discussed above,
 this mode
proceeds through the exchange of the superheavy Higgsino color triplet as can
be
seen in Fig. 5 [22,23].  Current data [20] gives the bound $\tau(p\rightarrow
{\bar \nu} K^+) > 1 \times 10 ^{32}$ yr $(90\%~ {\rm CL})$.  From Fig. 5,
one sees that the
amplitude for decay depends on $1/M_{H_3}$.  The current data then puts a bound
of $M_{H_3} \rs 1 \times 10^{16}~{\rm GeV}$ [24].  Future experiments expect an
increased sensitivity for Super Kamiokande of up to $\tau(p\rightarrow {\bar
\nu}K^+) < 2 \times 10^{33}$ yr [21], and for ICARUS of up to
$\tau(p\rightarrow
 {\bar
\nu}K^+) < 5 \times 10^{33}$ yr [25].  Thus the GUT model of Sec. 6, where
 $M_{H_3} < 2M_V$, would predict that if the $p\rightarrow e^+\pi^0$ mode
at future experiments were
observed, the $p\rightarrow {\bar \nu}K^+$ should be seen very copiously as
then
 $M_{H_3}$
would be less than $1.2  \times 10^{16}~{\rm GeV}$.
{}~\vskip 2.2truein
\noindent
Fig. 5.  Example of diagram contributing to the decay $p\rightarrow {\bar
\nu}_{\mu}K^+$.  There are additional diagrams with ${\bar \nu}_{\tau}$ and
${\bar\nu}_e$ final state.
CKM matrix elements appear at the $\tilde W$ vertices.
\smallskip
The p$\rightarrow {\bar \nu}K^+$ amplitude depends not only on $M_{H_3}$ but
also in a detailed way, on the SUSY masses of the particles in the loop of
Fig. 5 [23].  Since as discussed in Sec. 5, these masses are functions of the
 basic
parameters, which we may choose to be $m_0$, $m_{\tilde g} =
[\alpha_3(m_{\tilde g})/\alpha_G] m_{1/2}, A_t$ (The t-quark A parameter
at the electroweak scale) and tan $\beta$, the current bounds on p-decay
give rise to bounds in this parameter space.  If we restrict $M_{H_3} < 2
\times
10^{17}{\rm GeV}$ (which keeps $M_{H_3}/M_{P\ell} < 1/10$ and is what is
implied
 by
the GUT model of Sec. 6)  one finds the restrictions tan $\beta\l 8,
\mid A_t / m_0 \mid \l 2$ and in most of the parameter space $m_0
> m_{\tilde g}$.  Fig. 6 [26] shows what can be expected from future proton
 decay
experiments.  Thus if we require $m_0 \leq 1~ {\rm TeV}$ (to prevent
excessive fine
tuning), we see that ICARUS should detect $p\rightarrow {\bar \nu}K^+$ proton
decay for even the largest value of $M_{H_3}$ considered here (and Super
Kamiokande
should similarly detect this mode for $m_0\leq 950~{\rm GeV}$) if $m_{\tilde
W_1
} >
100~ {\rm GeV}$.  Thus if these experiments do not see proton decay,
then $m_{\tilde W_1} < 100~ {\rm GeV}$,
and hence the light Wino should be observable at LEP 200.  In
either case, $m_{\tilde W_1} < 100 ~{\rm GeV}$ or
$m_{\tilde W_1} > 100 ~{\rm GeV}$ these
models with $SU(5)$ type proton decay imply that a signal of supersymmetry
 should be
observed, and this could occur prior to the turning on of the LHC or SSC.

{}~\vskip 3.2truein
\noindent
Fig. 6.  Maximum value of $\tau(p\rightarrow {\bar \nu}K^+)$ for $m_t = 150~
{\rm GeV}$, $\mu < 0$ subject to the constraint $m_{\tilde W_1} > 100~ {\rm
GeV}
$.  The
dash-dot curve is for $M_{H_3} = 2 \times 10^{17}~{\rm GeV}$.  The dashed curve
for
$M_{H_3} = 1.2 \times 10 ^{17}~{\rm GeV}$, and the solid curve for $M_{H_3} = 6
\times 10
^{16}~{\rm GeV}$.  The horizontal upper and lower lines are the bounds of
ICARUS
and Super Kamiokande.

\medskip
\noindent
9.  CONCLUSIONS
\smallskip

Supersymmetry represents a natural way of solving the gauge hierarchy
problem.  Local supersymmetry, i.e. supergravity, supplies a formal structure
for treating supersymmetric grand unified models which allow for a consistent
treatment of spontaneous breaking of supersymmetry.  The supergravity GUT
models have a large amount of predictive ability in that the 32 SUSY particle
masses are determined from only five parameters.  One set of mass relations
which holds in several models over most of the parameter space is the
following scaling relations [24, 27]:

$$2m_{{\tilde Z}_1} \cong m_{{\tilde W}_1}\cong m_{{\tilde Z}_2}\eqno(28)$$

$$m_{{\tilde W}_2}\cong m_{{\tilde Z}_3}
\cong m_{{\tilde Z}_4} >> m_{{\tilde Z}_1}\eqno(29)$$

$$m_{{\tilde W}_1} \simeq {1\over 3} m_{\tilde g} ~{\rm for}~ \mu < 0;
m_{{\tilde W}_1} \simeq {1\over 4}~m_{\tilde g}~ {\rm for}~ \mu > 0\eqno(30)$$

\noindent
and

$$m_{H^0} \cong m_A \cong m_{H^\pm} >> m_h\eqno(31)$$

\noindent
These relations are actually the remnants of the gauge hierarchy problem.  Thus
in most of the allowed parameter space one has $m_0^2$, $m_{\tilde g}^2 >>
M_Z^2$ (which occurs already when $m_0$, $m_{\tilde g} \rs (2-3) M_Z$).  In
the radiative breaking equations, this usually means then that $\mu^2 >>
M_Z^2$ to guarantee enough cancellation so that the r.h.s. of the first
equation in Eq. (24) correctly adds up to only ${1\over 2}M^2_Z$.  One can then
check that Eqs. (28) - (31) are a consequence of $\mu^2 >> M_Z^2$ etc.  A
verificaiton of Eqs. (28) - (31) would be strong support of supergravity GUT
models as they depend strongly on how the structure of the theory at the GUT
scale accomplishes $SU(2) \times U(1)$ breaking at the electroweak scale.
\smallskip
Finally, we should like to stress that in spite of the ability of supergravity
GUT models to make testable predictions such as the ones discussed above, even
if it is a valid idea, it must still be viewed as an approximate effective
theory holding at scales below $M_G$.  The closeness of $M_G$ to $M_{P\ell}$,
i.e. $M_G/M_{P\ell} \simeq 1/10 -1/100$, implies then that the theory may
possess $\approx (1-10) \%$ errors in its predictions, and precision
experiments on the validity of these models could conceivably yield information
on the nature of Planck scale physics.

\medskip
\noindent
ACKNOWLEDGEMENTS
\smallskip
This work was supported in part by the National Science Foundation Grants Nos.
PHY-916593 and PHY-93-06906.  One of us (R.A.) would like to thank the
Department of Theoretical Physics, Oxford University for its kind hospitality
during the writing of this report.
\medskip
\noindent

References and Footnotes
\smallskip
\item{[1]} P. Langacker, Proc. PASCOS 90-Symposium, Eds. P. Nath and S.
Reucroft (World Scientific, Singapore 1990); J. Ellis, S. Kelley and D. V.
Nanopoulos, Phys. Lett. \underbar{249B}, 441 (1990); \underbar{B260}, 131,
(1991); V. Amaldi, W. De Boer and H. F\"urstenau, Phys. Lett.
\underbar{260B}, 447 (1991); F. Anselmo, L. Cifarelli, A. Peterman and A.
Zichichi, Nuov. Cim. \underbar{104A}, 1817 (1991); \underbar{115A}, 581
(1992).

\item{[2]}  H. Bethke, XXVI Conference on High Energy Physics, Dallas, 1992,
Ed.
 J. Sanford,
AIP Conf. Proc. No. 272 (1993); G. Altarelli, talk at Europhysics Conference on
High Energy
Physics, Marseille, 1993.

\item{[3]}  G. L. Kane et al, Phys. Rev. Lett. \underbar{70}, 2686 (1993).

\item {[4]}  J. F. Gunion, H. E. Haber, G. Kane and S. Dawson, ``The Higgs
Hunter's Guide" (Addison-Wesley, Reading, MA, 1990) [Erratum:  SCIPP-92/58
(1992)].

\item {[5]}  D.  Freedman, S. Ferrara and P. van Nieuwenhuizen, Phys. Rev.
\underbar{D13}, 3214 (1976); S. Deser and B. Zumino, Phys. Lett.
\underbar{B62}, 335 (1976).

\item {[6]}  E. Cremmer, S. Ferrara, L. Girardello and A. van Proeyen, Phys.
Lett. \underbar{116B}, 231 (1982); Nucl. Phys.\underbar{B212}, 413 (1983).

\item {[7]} A. H. Chamseddine, R. Arnowitt and P. Nath, Phys. Rev. Lett.
\underbar{29}, 970 (1982).

\item {[8]}  P. Nath, R. Arnowitt and A. H. Chamseddine, ``Applied N=1
Supergravity"  (World Scientific, Singapore, 1984).

\item {[9]}  E. Witten and J. Bagger, Nucl. Phys. \underbar{B222}, 125 (1983).

\item{[9a]} B. de Wit and D.Z. Freedman, Phys. Rev. Lett. 35, 827 (1975).

\item {[10]}  J. Polonyi, Univ. of Budapest Rep. No. KFKI-1977-93 (1977).

\item {[11]}  H. P. Nilles, Phys. Lett. \underbar{B115}, 193 (1981); S.
Ferrara, L. Girardello and H. P. Nilles, Phys. Lett. \underbar{B125}, 457
(1983).

\item {[12]}  H. P. Nilles, Phys. Rep. \underbar{110}, 1 (1984).

\item {[13]}  K. Inoue et. al., Prog. Theor. Phys. \underbar{68}, 927 (1982);
L. Iba\~nez and G. G. Ross, Phys. Lett. \underbar{B110}, 227 (1982); L.
Alvarez-Gaum\'e, J. Polchinski and M. B. Wise, Nucl. Phys. \underbar{B250}, 495
(1983); J. Ellis, J. Hagelin, D. V. Nanopoulos and K. Tamvakis, Phys. Lett.
\underbar{B125}, 2275(1983); L. E. Iba\~nez and C. Lopez, Phys. Lett.
\underbar{B128}, 54
(1983); Nucl. Phys. \underbar{B233}, 545 (1984); L. E. Iba\`nez, C. Lopez and
C.
 Mu\~nos, Nucl.
Phys. \underbar{B256}, 218 (1985); J. Ellis and F. Zwirner, Nucl. Phys.
\underbar{B388}, 317 (1990).

\item {[14]}  E. Witten, Nucl. Phys. \underbar{B177}, 477 (1981);
\underbar{B185} 513 (1981), S. Dimopoulos and H. Georgi, Nucl. Phys.
\underbar{B193}, 150 (1981); N. Sakai, Zeit. f. Phys. \underbar{C11}, 153
(1981).

\item {[15]} K. Inoue, A. Kakuto and T. Tankano, Prog. Theor. Phys.
\underbar{75}, 664 (1986); A. Anselm and A. Johansen, Phys. Lett.
\underbar{B200}, 331 (1988); A. Anselm, Sov. Phys. JETP. \underbar{67}, 663
(1988);R. Barbieri, G. Dvali and A. Strumia, Nucl. Phys. \underbar{B391}, 487
(1993).

\item {[16]}  R. Barbieri and L. J. Hall, Phys. Rev. Lett. \underbar{68}, 752
(1992); A. Faraggi, B. Grinstein and S. Meshkov, Phys. Rev. \underbar{D47},
5018 (1993); L. Hall and U. Sarid, Phys. Rev. Lett. \underbar{70}, 2673
(1993).

\item{[17]}  D. Ring and S. Urano, unpublished (1992).

\item {[18]} While the world average for $\alpha_3(M_Z)$ is $0.118 \pm 0.007$
[2], the $1\sigma$ value on the high side, $\alpha_3(M_Z) = 0.125$, ${\rm is~
already}~
2.6\sigma$ above the low energy deep inelastic scattering result of
$\alpha_3(M_Z) =0.112 \pm 0.005$.  Thus one can not really go much above the
$\alpha_3(M_Z) = 0.125$ without assuming some systematic error exists in
the low energy measurements of $\alpha_3$.

\item {[19]} P. Langacker and N. Polonsky, Phys. Rev. \underbar{D47}, 4028
(1993).

\item {[20]} Particle Data Group, Phys. Rev. \underbar{D45}, Part 2 (June,
1992).

\item {[21]} Y. Totsuka, XXIV Conf. on High Energy Physics, Munich, 1988,
Eds. R. Kotthaus and J. H. Kuhn (Springer Verlag, Berlin, Heidelberg, 1989).

\item {[22]}  S. Weinberg, Phys. Rev. \underbar{D26}, 287 (1982); N. Sakai
and T. Yanagida, Nucl. Phys. \underbar{B197}, 533 (1982); S. Dimopoulos, S.
Raby and F. Wilczek, Phys. Lett. \underbar{B112}, 133 (1982); S. Chadha and
 M. Daniels, Nucl.
Phys. \underbar{B229}, 105 (1983);
B. A. Campbell, J. Ellis, and D. V. Nanopoulos, Phys. Lett.
\underbar{B141}, 224 (1984).

\item {[23]}  R. Arnowitt, A. H. Chamseddine and P. Nath, Phys. Lett.
\underbar{B156}, 215 (1985); P. Nath, R. Arnowitt and A. H. Chamseddine,
Phys. Rev. \underbar{D32}, 2348 (1985).

\item {[24]}  R. Arnowitt and P. Nath, Phys. Rev. Lett. \underbar{69}, 725
(1992).

\item {[25]} ICARUS Detector Group, Int. Symposium on Neutrino Astrophysics,
Takayama, 1992.

\item {[26]} R. Arnowitt and P. Nath, CTP-TAMU-32/93-NUB-TH-3066/93-SSCL
Preprint-440 (1993).

\item {[27]}  J. L. Lopez, D. V.Nanopoulos and A. Zichichi, CERN-TH 6667
(1993).
\vfill\eject\bye